# Ag-Au alloys BCS-like Superconductors?

Surender Singh, Subhamoy Char, and Dasari L. V. K. Prasad[*]
*Department of Chemistry, Indian Institute of Technology, Kanpur 208016, India*
(Dated: December 21, 2018)

Prompted by the recent report on the evidence for superconductivity at ambient temperature and pressure in nanostructures of silver particles embedded into a gold matrix [arXiv:1807.08572], we have exploited first principles materials discovery approaches to predict superconductivity in the 3D bulk crystals and 2D slabs of Ag-Au binary alloys at 1 atm pressure within the phonon-mediated BCS-like pairing mechanism. In calculations, it turns out that, the estimated superconducting transition temperatures of the ensued stable and metastable Ag-Au alloys resulted in $T_c$ as low as one mK. Whereas similar calculations for the known superconducting intermetallic compounds consisting of gold, Laves $Au_2Bi$ and A15 $Nb_3Au$ predict $T_c \approx 3.6$ K and 10.1 K, respectively, corroborate with experiments. And, the hitherto unknown silver analogues, $Ag_2Bi$ and $Nb_3Ag$ are also found to be superconducting at 6.1 K and 10.8 K, respectively. Furthermore, we show that, elemental Au in its metastable 9*R*, and *hcp* phases superconduct at $T_c \approx 1$ mK; but not Ag, in these hexagonal lattices.

*Introduction*: The discovery of superconductivity in mercury below 4.12 K by Kamerlingh Onnes[1] has led to a resurgent research interest in the development of materials that superconduct at ambient temperature and pressure (T=298 K and P = 1 atm).[2,3] In light of this, over the period of more than a century, a vast number of diverse materials, starting from the structurally simple elemental solids to complex cuprate extended networks were identified in experiments and as well as in theory with increasing superconducting transitions temperatures ($T_c$).[4,5,6,7,8,9] While the lowest $T_c$ reported so far is for Li at 0.4 mK[10,11] and Bi at 0.53 mK;[12] the highest $T_c$ is 133 K, recorded in cuprates (Hg-1223),[13] all at 1 atm. The other important superconducting class of compounds are A15,[14,15] $MgB_2$,[16] Iron-Pnictides,[17] and compressed hydrogen–and–the chemically precompressed hydrogen-rich alloys,[18,19,20,21,22] in particular the recent H-S family of compounds, the later have been predicted initially in computations[23,24] and soon after verified in experiments with an incredible $T_c$ as high as 203 K under diamond anvil pressures of about 150 GPa,[25] (see also ref 26,27). Evidently, these conjunctive findings continued to inculcate in searching further for high $T_c$ materials – be it in theory or in experiments, to eventually find *the material* that superconduct at room temperature.

To this effect, in recent experiments, Thapa and Pandey observed signals for room temperature superconductivity (RTS) at ambient pressure in Au-rich Ag-Au alloy nanostructures.[28] It is interesting to note that, in such an RTS, neither pure components (Ag and Au) in their native states (*fcc*), is alone known to be superconducting, down to the lowest reachable temperatures measured so far for these elements.[29,30,31] *Is it a somewhat frivolous*

---

[*] To whom correspondence should be addressed. Email: dprasad@iitk.ac.in

*correlation?* Analysis of the materials data on superconductors indicates that numerous compounds are known to be superconducting (at 1 atm), but not their constituent elements.[32,33,34,35] For example, $MgB_2$ is a bulk superconductor with $T_c \approx 39$ K, where the elements Mg and B are not superconductors in their ground states at 1 atm.

Superconductivity in Au-compounds is not uncommon either. Soon after the discovery of superconductivity in Hg, one of the first alloys measured for superconductivity by the Leiden team was Hg-Au ($T_c \sim 4$ K, liquid-helium temperatures).[36] Subsequently, superconductivity has been observed in a fistful of gold alloys.[37,38] Among all, the well-characterized binary superconducting compounds are: the pretty Laves $Au_2Bi$[39] and A15 $Nb_3Au$[40] intermetallic alloys with $T_c$ ~1.84 K and ~11.5 K, respectively. Note the increased transition temperatures of the alloys in reference to their elements Bi (0.53 mK)[12] and Nb (9.25 K)[4]. As will be seen, in our density functional theory (DFT) based electron–phonon calculations coupled to modified Bardeen-Cooper-Schrieffer (BCS) theory, the results not only indicate that these gold consisting alloys are superconductors, but also reveal superconductivity in the corresponding hitherto unknown silver analogues ($Ag_2Bi$ and $Nb_3Ag$). In view of this, it may be appropriate and imperative to ask: Are the Ag-Au alloys BCS-like superconductors?

As cited earlier, although both the *fcc* Ag and Au elements are not superconductors, it was suggested in a recent article by Baskaran[41] that superconductivity may be confined at the interfaces of Ag-Au alloys due to the perturbation driven quasi 2D structural reconstructions (*fcc* to 9*R* also called as *hR*9 phase in hexagonal setting) and thereby the emergent local quasi 2D Fermi surfaces. It was also proposed in literature that the epitaxial thin films of Au(111) and Ag(111) surfaces are ideal for proximity induced superconductivity.[42] For example, in a recent study on V/Au bilayer heterostructure, it has been successfully demonstrated that the Au(111) films superconduct at $T_c \sim 3.9$ K.[43]

In this article, encouraged by these intuitive structure-property correlations, emerging properties, and the evidence[28] and also the absence (Ogale et al.)[44] of RTS in Ag-Au alloy nanostructures, we have theoretically investigated structures of Ag-Au bulk alloys and 2D slabs for any plausible superconductivity, at ambient pressure. As the alloy structures, the atomic ordering in the crystal lattices are not well defined in the experimental findings of Pandey et al. and Ogale et al., here we predict stable and metastable Au-rich crystal structures of Ag-Au bulk alloys at various compositions ($Ag_xAu_{1-x}$, where x=0.5 to 0.17) and the 2D slabs are modeled based on the *fcc* (111) surface structure. Thereby, for the best stable and other interesting structural candidates, the superconductivity has been calculated. Indeed, in the structures we predicted, as will be shown below, that the 9*R*-like Ag-Au alloy structures are <u>only</u> found to be superconductors, but with $T_c$ not more than a mK. And, we also show that the Ag and Au elements are metastable in hexagonal lattices, *hcp, hP*4, and 9*R* phases, in which only the elemental gold 9*R* and *hcp* phases displayed superconducting transition temperature; it is estimated to be in the order of 1 mK. To our surprise – or maybe not –, as discussed in detail below, we did not find increased $T_c$ in any of the Ag-Au bulk alloys and the 2D slabs that we have calculated.

*Structure-searches*: For crystal structure prediction, probable $Ag_xAu_{1-x}$ stoichiometric compositions are chosen based on the superconductivity studies presented in reference 28 (see Figure 4a, Figure S2, S3, and S8 in reference 27). In particular for x ≈ 0.2 ($AgAu_4$) and 0.17 ($AgAu_5$), the $T_c$ has been reported as 320 K and 240 K, respectively. And, as a general trend, it was shown there that the $T_c$ decreases with the increased Au concentration in $Ag_xAu_{1-x}$ alloy nanostructures. Therefore, we have systematically investigated the structure searches for x = 0.5, 0.33, 0.25, 0.2, and 0.17 – the Au-rich side of the Ag-Au alloy phase diagram. Even with this information at hand, predicting a structure for a given composition is computationally an intensive task. Therefore, the Ag-Au alloy crystal structures are initially predicted based on the *Strukturbericht* crystal structure-type classification, and thereby with the evolutionary crystal structure algorithms[45] coupled with DFT calculations.[46] This procedure is adapted to quicken the process of predicting crystal structures, identify any missing stable and metastable structures, and as well as to incorporate structures with large Z-values (Z>4). This approach[47] and the crystal data with statistics will be published elsewhere. For each composition, we have considered 1, 2, 3, and 4 formula units (Z) per unit cell. The predicted list of the crystal structures of $Ag_xAu_{1-x}$ binary alloys is shown in a scatter plot in Figure 1. More information about the computational details is provided in the Supplementary Material.

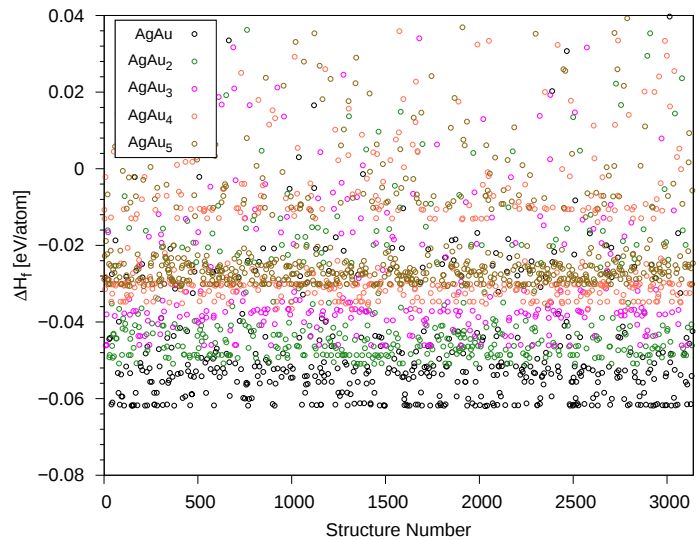

**Figure 1**. Evolution of the crystal structures predicted for Ag-Au binary alloys. The calculated enthalpies of formation of all the structures (below 0.04 eV/atom) are plotted over the structure number. The ordinate cut-off is chosen only to visually appreciate the composition differences shown on the plot. The enthalpies are calculated from the ground state Ag and Au *fcc* structures. The structures examined are color-coded according to the alloy composition, as shown in the inset at upper-left.

*Phase stability*: In general, the thermodynamic stability of the alloys is investigated by constructing a convex hull – a tie-line graph of the enthalpy of formation ($\Delta H_f$) verses the alloy composition. If the alloys are with negative enthalpy of formation, and the enthalpy of decomposition of these into any other neighboring compositions is positive, then the alloy is thermodynamically stable. Figure 2a shows the Ag-Au tie-line of the enthalpies of formation for the selective candidate structures with $\Delta H_f < 0.04$ eV/atom at each Ag-Au composition examined. In Figure 2b, four lowest energy structures for each composition and "interesting" competitive metastable phases (< 0.04 eV/atom) are depicted. They are 58 structures in total, all are dynamically stable and for all of which the superconducting transition temperatures are estimated. It is interesting to note that in the best four low energy structures in each Ag-Au composition there is at-least one structure reported in *Strukturbericht,* the others in the list are found to be much higher in energy with positive heats of formations or just below the reference-line, as shown in Figure 2a.

As can been seen from the Figure 2b, for all the ordered Ag-Au alloy phases, the $\Delta H_f$ are significantly negative, which shows complete miscibility (or solid solubility) across the phase diagram. The AgAu (1:1) composition has the most negative enthalpy of formation. This is in consistent with the previous experimental and theoretical findings.[48,49,50,51,52,53,54,55] In our crystal structure prediction, in addition to the Ag-Au known phases in literature, we have identified a number of novel structures, in particular for $AgAu_2$ and $AgAu_4$ and $AgAu_5$ compositions. A detailed description of the relative energy comparisons for the best four structures in each composition and other interesting phases are provided in the Supplementary Material.[56] We have incorporated as many structures as computationally feasible for us to investigate superconductivity.

Since the Ag-Au alloys form stable solid solutions, finite temperature contributions are critical to develop accurate phase diagrams. The temperature corrected Ag-Au convex hull is shown in Figure 2c. The free energies are calculated by incorporating the vibrational entropic contributions. At each designated temperature, the corresponding ground state alloy structures have been considered to plot the free energy tie-line. Stability range of the $Ag_xAu_{1-x}$ binary alloys with respect to temperature is shown in Figure 2d. As can be seen, at finite temperatures, the $AgAu_2$, $AgAu_3$, and $AgAu_4$ do undergo temperature induced structural phase transformations. Nevertheless, all persist with negative heats of formation. As the temperature increases the free energy tie-lines are tend to be more shallow, vindicating the isomorphous-like solid solution behavior of Ag-Au alloys.

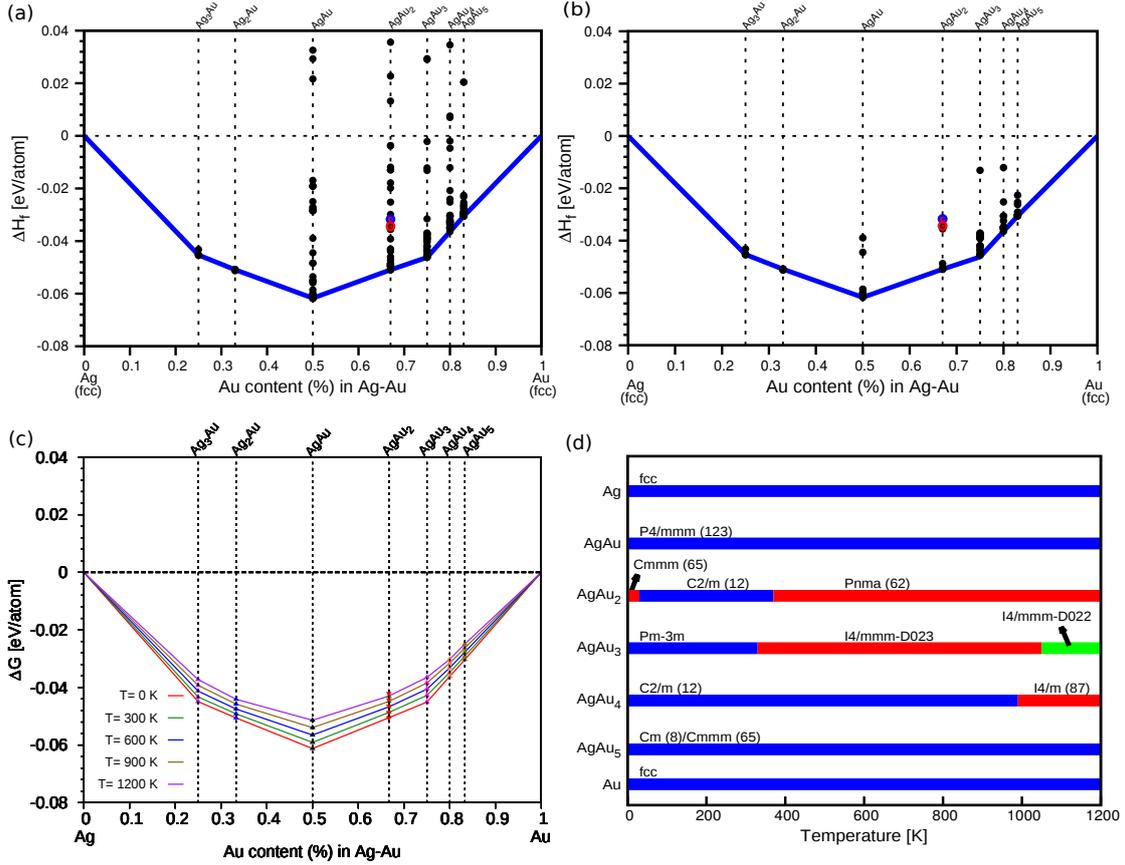

Figure 2. (a and b) static, (c) temperature corrected convex hull diagrams, and (d) temperature-composition phase diagram of the Ag-Au alloy system. In (a), (b) and (c), solid lines denote the convex hulls and the symbols represent the enthalpies of individual phases. The compositions on the solid line are thermodynamically stable, while those off the solid lines are metastable; they may decompose into other near by alloys on the phase diagram. In (a) and (b), the AgAu$_2$ metastable phases marked in red and blue are superconductors (see text for details). The compositions are indicated at the top abscissa. In (d) the compositions are indicated at the left ordinate. The temperature ranges in which the structures undergo phase transformations are color-coded and each phase labeled by its Hermann–Mauguin crystallographic space group notation. All the enthalpies and free energies are calculated relative to the ground state elemental Ag (*fcc*) and Au (*fcc*) crystal structures.

*Crystal Structures*: Figure 3 illustrates the structures of the most stable Au-rich Ag-Au alloys (the structures on the solid lines at T → 0 K in Figure 2a, 2b, and 2c): AgAu (*P4/mmm*), AgAu$_2$ (*Cmmm*), AgAu$_3$ (*Pm-3m*), AgAu$_4$ (*C2/m*), and AgAu$_5$ (*Cm*) phases. The structures are three-dimensionally interconnected by forming coordinated polyhedral networks with Ag-Ag, Au-Au and Ag-Au internuclear separations (2.92 Å) much comparable to the calculated (2.93 Å) and experimental values in *fcc* Ag (2.89 Å) and Au (2.88 Å) elemental solids and cubic AgAu (2.88 Å) disordered alloy. The distance similarities can be attributed due to the fact that Ag and Au share common crystal packing with similar atomic radii. Interestingly, almost all of the stable Ag-Au alloy structures predicted here consist of cuboctahedron as a common structural motif, shown in Figure 4. This essentially indicates that the ground state Ag-Au alloy structures can be classified as *fcc* variants. Could it be that the crystal structures predicted here are the bulk analogues of the nanostructures synthesized in reference 28 and 44? See Supplementary Material for a detailed description of the structural parameters and their comparison with corresponding experimental values.

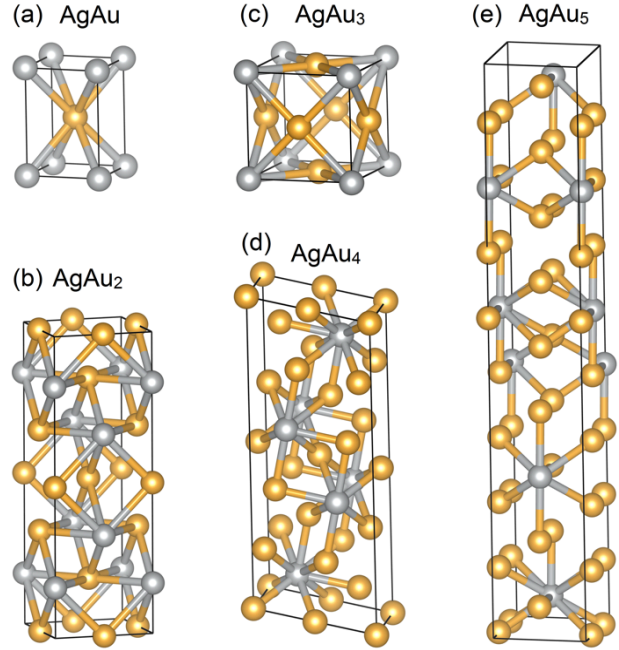

Figure 3. Perspective crystalline views of the predicted most stable Ag-Au alloys with space groups (a) AgAu (*P4/mmm*), (b) AgAu$_2$ (*Cmmm*), (c) AgAu$_3$ (*Pm-3m*), (d) AgAu$_4$ (*C2/m*), and (e) AgAu$_5$ (*Cm*). Ag atoms are depicted as grayish-silver spheres and Au atoms are yellowish-gold. The internuclear separation between the nearest neighbor Ag and Au are shown as bicolor sticks, for a cut-off distance of 3 Å.

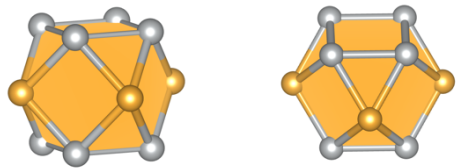

Figure 4. Cuboctahedron in Ag-Au alloys shown in (right) skew and (left) orthogonal projections centered on a triangular face.

*Superconductivity*: Electron-phonon coupling (EPC) calculations for the best stable structures shown in Figure 3 yield $\lambda$ values as low as 0.16 to 0.17 and the logarithmic average frequency $\omega_{\log}$ is 109 K to 120 K. The electronic densities of states near the Fermi Energy, N(E$_F$) are not great either (see Table in Supplementary Material). Superconducting transition temperatures ($T_c$) are estimated by applying the Allen-Dynes modified McMillan equation with a choice of Coulomb potential $\mu^*$=0.1. For such small values of $\lambda$ and $\omega_{\log}$, the estimated $T_c$ are found to be zero.

As one of the prime elements of this study is to search for superconductivity in Ag-Au system, the next best stable or interesting metastable structures depicted in Figure 2b have been explored for superconductivity. While as such there is no hard written rule that would direct which phase is more interesting (other than the low energy structures) to study for superconductivity in Ag-Au alloys, we have considered *fcc* and *hcp* like phases for which room temperature superconductivity has been proposed in experiments[28] and in theory,[41] respectively. The results are not encouraging, superconductivity has not been found above to 1 mK in any of the studied Ag-Au alloys, including 2D slabs. For three hexagonal AgAu$_2$ phases shown in Figure 5, a $T_c$ of one mK is estimated. This may be attributed to a slightly improved $\lambda$ of 0.23. The $\lambda$, $\omega_{log}$, N(E$_F$), and $T_c$ for each stable and metastable Ag-Au alloys are reported in Supplementary Material. As can be inferred from this Table, according to BCS or its modified versions of phonon-mediated theories of superconductivity, the factors responsible for the occurrence of possible superconductivity are absent in Ag-Au alloys.

The metastable AgAu$_2$ hexagonal structures are interesting. The most stable structure is of CdI$_2$ prototype (*R-3m*) with a complex arrangement of hexagonal layers (see Figure 5a). The other two are with staking sequence of *α*-Sm (or Li 9*R*) prototype. But they are differed by inversion symmetry (*R3m* and verses *R-3m*). It is not uncommon to have the hexagonal staking sequences in elemental Ag and Au. Recently, it has been shown in many crystal phase controlled synthesis of noble metal nanostructures that the hexagonal phases 2H and 4H can be made.[57,58] In our calculations, the hexagonal *h*P2, *h*P4, and *h*R9 phases of Ag and Au are found to be metastable (they are dynamically stable as well). With respect to *fcc,* the hexagonal phases are higher in energy in the order of *h*P2>*h*R9>*h*P4>*fcc*. The relative energies (relativistic effects are included as well) are reported in the supplementary material. The estimated $T_c$ for Au 9*R*, and *hcp* phases is 1 mk ($\mu^*$=0.1). In other Ag and Au phases studied here, the $T_c$ is found to be zero. The calculated superconducting transition temperatures ($\mu^*$=0.1) for the gold consisting alloys, Laves Au$_2$Bi and A15 Nb$_3$Au are in good agreement with the experimental values.

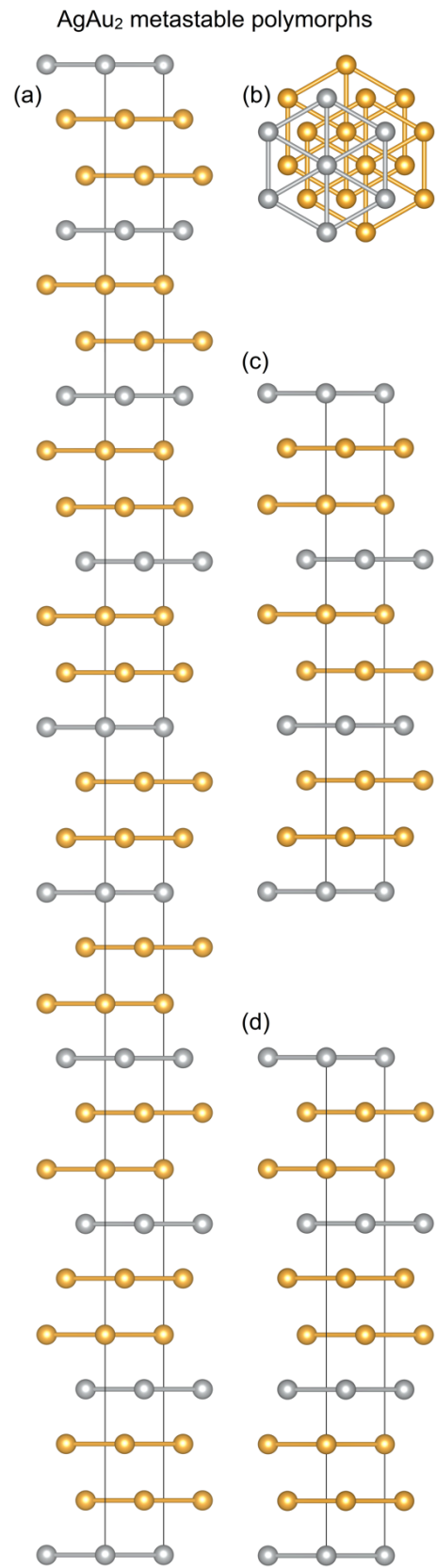

Figure 5. Perspective crystalline views of the metastable AgAu$_2$ polymorphs (a) *R*-3*m* (CdI$_2$ type), (b) top view of (a), (c) *R*3*m*, and (d) *R*-3*m* ($\alpha$-Sm type).

*Conclusion:* In our calculations, within the phonon-mediated superconducting mechanism, we do not find signature of superconductivity above 1 mK. The chief reasons could be that it may well have to do with the weak electron-phonon coupling constant that persisted in equal magnitudes irrespective of the structure and composition of the Ag-Au alloys studied here.

The authors acknowledge the Computer Center High Performance Computing facility at Indian Institute of Technology Kanpur and HPC/RNJJ for computational resources. S.S. an S. C. thanks the CSIR, India for a SRF research fellowship. D.L.V.K.P. acknowledges financial support through Initiation Grant No. IITK/CHM/20130116.

Supplementary Material can be downloaded from the link below:
http://home.iitk.ac.in/~dprasad/papers/AgAu_Supplementary _Material.tar.gz

References

[1] H. K. Onnes, "The Resistance of Pure Mercury at Helium Temperatures", Commun. Phys. Lab. Univ. Leiden, **119b** (Feb. 1911), reprinted in Proc. K. Ned. Akad. Wet. **13**, 1107 (1911).

[2] V. L. Ginzburg, "Nobel Lecture: On superconductivity and superfluidity (what I have and have not managed to do) as well as on the "physical minimum" at the beginning of the XXI century", Rev. Mod. Phys. **76**, 981 (2004).

[3] M. L. Cohen, "Essay: Fifty Years of Condensed Matter Physics", Phys. Rev. Lett. **101**, 250001 (2008).

[4] B. T. Matthias, T. H. Geballe, and V. B. Compton, "Superconductivity", Rev. Mod. Phys. 35, 1 (1963) and Erratum Rev. Mod. Phys. 35, 414 (1963).

[5] J. G. Bednorz and K. A. Müller, "Possible highTc superconductivity in the Ba−La−Cu−O system", Z. Phys. B, **64**, 189 (1986).

[6] N. W. Ashcroft, "Hydrogen Dominant Metallic Alloys: High Temperature Superconductors?", Phys. Rev. Lett. **92**, 187002 (2004).

[7] B. Janko, G. W. Crabtree, and W.-K. Kwok (Edited), "Room Temperature Superconductivity", Physica C **468**, 97 (2008).

[8] J. E. Hirsch, M. B. Maple, and F. Marsiglio (Edited), "Superconducting Materials: Conventional, Unconventional and Undetermined", Physica C **514**, 1 (2015).

[9] Lev. P. Gor'kov and V. Z. Kresin, Colloquium: High pressure and road to room temperature superconductivity, Rev. Mod. Phys. **90**, 011001 (2018) and references there in.

[10] J. Tuoriniemi, K. Juntunen-Nurmilaukas, J. Uusvuori, E. Pentti, A. Salmela, and A. Sebedash, "Supercoductivity in lithium below 0.4 millikelvin at ambient pressure", Nature, **447** 187 (2007).

[11] Polycrystalline rhodium samples were found to be superconducting at ≈ 0.3 mK. It may be that the $T_c$ is below Li. However, it was argued that the samples were magnetically contaminated. See, Ch. Buchal, F. Pobell, R. M. Mueller, M. Kubota, and J. R. Owers-Bradley, "Superconductivity of Rhodium at Ultralow Temperatures", Phys. Rev. Lett. **50**, 64 (1983) and Ch. J. Raub, "Superconductivity of the platinum metals and their alloys", Platinum Metals Rev. **28**, 63 (1984).

[12] O. Prakash, A. Kumar, A. Thamizhavel, and S. Ramakrishnan, "Evidence for bulk superconductivity in pure bismuth single crystals at ambient pressure", Science **355**, 52 (2017).

[13] A. Schilling, M. Cantoni, J. D. Guo, and H. R. Ott, "Superconductivity above 130 K in the Hg–Ba–Ca–Cu–O system", Nature 363, 56–58 (1993).

[14] G. F. Hardy, J. K. Hulm, "The Superconductivity of Some Transition Metal Compounds", Phys. Rev. **93**, 1004 (1954).

[15] G. R. Stewart, "Superconductivity in the A15 structure", Physica C **514**, 28 (2015).

[16] J. Nagamatsu, N. Nakagawa, T. Muranaka, Y. Zenitani, and J. Akimitsu "Superconductivity at 39 K in magnesium diboride" Nature **410**, 63 (2001).

[17] Y. Kamihara, H. Hiramatsu, M. Hirano, R. Kawamura, H. Yanagi, T. Kamiya, and H. Hosono, "Iron-Based Layered Superconductor: LaOFeP", J. Am. Chem. Soc. **128**, 10012 (2006).

[18] E. Zurek, R. Hoffmann, N. W. Ashcroft, A. R. Oganov, and A. O. Lyakhov, "A little bit of lithium does a lot for hydrogen", Proc. Natl. Acad. Sci. USA **106**, 17640 (2009).

[19] G. Gao, A. R. Oganov, P. Li, Z. Li, H. Wang, T. Cui, Y. Ma, A. Bergara, A. O. Lyakhov, T. Iitaka, and G. Zou, "High-pressure crystal structures and superconductivity of Stannane ($SnH_4$)", Proc. Natl. Acad. Sci. U.S.A. **107**, 1317 (2010).

[20] H. Wang, J. S. Tse, K. Tanaka, T. Iitaka, and Y. Ma, "Superconductive sodalite-like clathrate calcium hydride at high pressures", Proc. Natl. Acad. Sci. U.S.A. **109**, 6463 (2012).

[21] H. Liu, I. I. Naumov, R. Hoffmann, N. W. Ashcroft, and R. J. Hemley, "Potential high-Tc superconducting lanthanum and yttrium hydrides at high pressure", Proc. Natl. Acad. Sci. U.S.A. **114,** 6990 (2017).

[22] T. Bi, N. Zarifi, T. Terpstra, and E. Zurek, "The Search for Superconductivity in High Pressure Hydrides", arXiv:1806.00163 and references therein.

[23] Y. Li, J. Hao, H. Liu, Y. Li, and Y. Ma, "The Metallization and Superconductivity of Dense Hydrogen Sulfide", J. Chem. Phys. **140**, 174712 (2014).

[24] D. Duan, Y. Liu, F. Tian, D. Li, X. Huang, Z. Zhao, H. Yu, B. Liu, W. Tian, and T. Cui "Pressure–Induced Metallization of Dense $(H_2S)_2H_2$ with high–Tc Superconductivity", Sci. Rep. **4**, 6968 (2014).

[25] A. P Drozdov, M. I. Eremets, I. A. Troyan, V. Ksenofontov, S. I. Shylin, "Conventional Superconductivity at 203 Kelvin at High Pressures in the Sulfur Hydride System", Nature **525**, 73 (2015). See also reference [**25**], wherein, it has been shown that $LaH_x$ superconducts at ~215 K under ~150 GPa.

[26] A. P. Drozdov, V. S. Minkov, S. P. Besedin, P. P. Kong, M. A. Kuzovnikov, D. A. Knyazev, and M. I. Eremets, "Superconductivity at 215 K in lanthanum hydride at high pressures", arXiv:1808.07039

[27] Maddury Somayazulu, Muhtar Ahart, Ajay K Mishra, Zachary M. Geballe, Maria Baldini, Yue Meng, Viktor V. Struzhkin, and Russell J. Hemley, arXiv:1808.07695

[28] D. K. Thapa and A. Pandey, "Evidence for Superconductivity at Ambient Temperature and Pressure in Nanostructures", arXiv:1807.08572

[29] R. F. Hoyt and A. C. Mota, "Superconductivity in α-phase alloys of Cu, Ag and Au", Solid State Commun. **18**, 139, (1976).

[30] K. Ôno, K. Asahi, N. Nishida, J. Ray, and H. Ishimoto, "Search for superconductivity in gold below 100 μK", Physica **107B**, 719 (1981).

[31] Ch. Buchal, R. M. Mueller, F. Pobell, M. Kubota, and H. R. Folle, "Superconductivity investigations of Au-In alloys and of Au at ultralow temperatures", Solid State Commun. **42**, 43 (1982). Here, in order to determine the $T_c$ of pure Au, the measured $T_c$ values of $Au_{1-x}In_x$ alloys have been extrapolated to x=0. In such estimates, it was found that gold becomes superconducting at T ~ 100 mK.

[32] B. T. Matthias, "Superconducting compounds of nonsuperconducting elements", Phys. Rev. **90**, 487 (1953).

[33] B. T. Matthias, "Transition temperatures of superconductors", Phys. Rev. **92**, 874 (1953).

[34] CRC Handbook of Chemistry and Physics, edited by D.R. Lide (CRC Press, 1995), pp. 12-90.

[35] B. W. Roberts, "Superconductive materials and some of their properties" General Electric Co. Research Lab., Schenectady, NY, 1963).

[36] H. K. Onnes, Commun. Phys. Lab. Univ. Leiden 133d, reprinted in Proc. K. Ned. Akad. Wet. **16**, 113 (1913).

[37] H. R. Khan, "Superconducting gold alloys", Gold Bull. **17**, 94 (1984).

[38] R. Flükiger and W. Klose, "Ac - Na. Landolt-Börnstein - Group III Condensed Matter (Numerical Data and Functional Relationships in Science and Technology)", Vol 21a. (Springer, Berlin, Heidelberg), pp 56-61.

[39] W.J. de Haas, F. Jurriaanse, "Die Supraleitfähigkeit des Gold-Wismuts" Naturwissenschaften, **19**, 706 (1931).

[40] E. A. Wood and B. T. Matthias, "The crystal structures of $Nb_3Au$ and $V_3Au$", Acta Cryst. **9**, 534 (1956).

[41] G. Baskaran, "Theory of Confined High $T_c$ Superconductivity in Monovalent Metals" arXiv:1808.02005

[42] A. C. Potter and P. A. Lee, "Topological superconductivity and Majorana fermions in metallic surface states", Phys. Rev. B **85**, 094516 (2012).

[43] P. Wei, F. Katmis, C-Z. Chang, and J. S. Moodera, "Induced Superconductivity and Engineered Josephson Tunneling Devices in Epitaxial (111)-Oriented Gold/Vanadium Heterostructures", Nano Lett. **16**, 2714 (2016).

[44] A. Biswas, S. Parmar, A. Jana, R. J. Chaudhary and S. Ogale, arXiv:1808.10699.

[45] D. C. Lonie and E. Zurek, "XtalOpt: An Open–Source Evolutionary Algorithm for Crystal Structure

Prediction", Comput. Phys. Commun. **182**, 372 (2011).

[46] G. Kresse and J. Furthmüller, "Efficient iterative schemes for ab initio total-energy calculations using a plane-wave basis set", Phys Rev B **54**, 11169 (1996).

[47] S. Char, M. S. Chauhan, S. Singh, A. Kumar, K. Yadav, and D.L.V.K. Prasad (unpublished).

[48] B. Predel, "Ag-Au (Silver-Gold). In: Madelung O. (eds) Ac-Au – Au-Zr" Landolt-Börnstein - Group IV Physical Chemistry (Numerical Data and Functional Relationships in Science and Technology), vol 5a. Springer, Berlin, Heidelberg

[49] L. Karmazin, "Accurate measurement of lattice parameters of Ag-Au solid solutions" Czech J. Phys. **19**, 634 (1969).

[50] Y. C. Venudhar, L. Iyengar, K. V. Krishna Rao, "Temperature dependence of the lattice parameter and the thermal expansion of Ag-Au (50 at. %) alloy by an X-ray method" J. Less-Common Met. **60**, 41 (1978).

[51] V. Ozoliņš, C. Wolverton, and Alex Zunger, "Cu-Au, Ag-Au, Cu-Ag, and Ni-Au intermetallics: First-principles study of temperature-composition phase diagrams and structures" Phys. Rev. B **57**, 6427 (1998).

[52] K. Terakura, T. Oguchi, T. Mohri, and K. Watanabe, "Electronic theory of the alloy phase stability of Cu-Ag, Cu-Au, and Ag-Au systems" Phys. Rev. B **35**, 2169 (1987).

[53] S. Curtarolo, D. Morgan, and G. Ceder, "Accuracy of ab initio methods in predicting the crystal structures of metals: A review of 80 binary alloys" Calphad **29**, 163 (2005).

[54] N. A. Zarkevich, Teck L. Tan, L.-L. Wang, and D. D. Johnson, "Low-energy antiphase boundaries, degenerate superstructures, and phase stability in frustrated fcc Ising model and Ag-Au alloys" Phys. Rev. B **77**, 144208 (2008).

[55] Atsuto Seko, Kazuki Shitara, and Isao Tanaka, "Efficient determination of alloy ground-state structures" Phys. Rev. B **90**, 174104 (2014).

[56] Since the evidence for superconductivity has been proposed in the systems of Au-rich Ag-Au nanostructures, we haven't explored the structure searches for the silver-rich side (for example, $Ag_3Au$ and $Ag_2Au$) of the Ag-Au phase diagram. The structural analogues of the best four stable structures of $AgAu_3$ and $AgAu_2$ are considered to construct the Ag-rich side of the convex hull.

[57] Z. Fan, X. Huang, Y. Chen, W. Huang, H. and Zhang, Nat Protoc. **12**, 2367 (2017).

[58] I. Chakraborty, D. Carvalho, S. N. Shirodkar, S. Kumar, S. Bhattacharyya, R. Banerjee, U. Waghmare, and P. Ayyub, J. Phys.: Condens. Matter **23** 325401 (2011).